\documentclass[prd,showpacs,amsmath,amssymb]{revtex4}
\usepackage{graphicx,color}
\begin{document}
\title{Study on contributions of hadronic loops
to decays of $J/\psi\rightarrow$ vector $+$ pseudoscalar mesons }

\author{Xiang Liu}\email{liuxiang726@mail.nankai.edu.cn}
\author{Xiao-Qiang Zeng}\email{zengxq@mail.nankai.edu.cn}
\author{Xue-Qian Li}\email{lixq@nankai.edu.cn}

\affiliation{Department of physics, Nankai University, Tianjin
300071, China}
\date{\today}

\vspace{0.8cm}

\begin{abstract}
In this work, we evaluate the contributions of the hadronic loops
to the amplitudes of $J/\psi\rightarrow PV$ where $P$ and $V$
denote light pseudoscalar and vector mesons respectively. By
fitting data of two well measured channels of $J/\psi\rightarrow
PV$, we obtain the contribution from the pure OZI process to the
amplitude which is expressed by a phenomenological quantity
$|\mathcal{G}^{PV}_{S}|$, and a parameter $\alpha$ existing in the
calculations of the contribution of hadronic loops. In terms of
$\alpha$ and $|\mathcal{G}^{PV}_{S}|$, we calculate the branching
ratios of other channels and get results which are reasonably
consistent with data. Our results show that the contributions from
the hadronic loops are of the same order of magnitude as that from
the OZI processes and the interference between the two
contributions are destructive. The picture can be applied to study
other channels such as PP or VV of decays of $J/\psi$.

\end{abstract}  \maketitle

\section{introduction}

By the commonly accepted point of view, narrowness of  the
$J/\psi$ resonance is interpreted by the OZI rule\cite{OZI}.
Namely, $J/\psi$ dissolves into three gluons which would then
hadronize into measurable hadrons and these processes are OZI
suppressed.  In Fig.1, the three-gluon process diagrams are shown
and they are single-OZI-suppressed (SOZI) or double-OZI (DOZI)
suppressed. Generally one can ignore the contributions from DOZI.
\begin{figure}[htb]
\begin{center}
\begin{tabular}{ccccc}
\scalebox{0.7}{\includegraphics{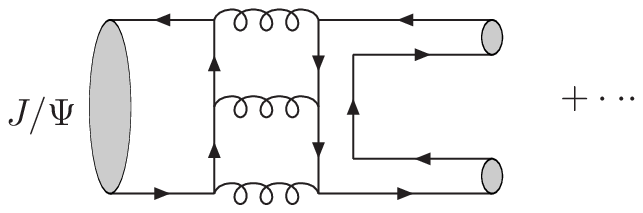}}&&\scalebox{0.7}{\includegraphics{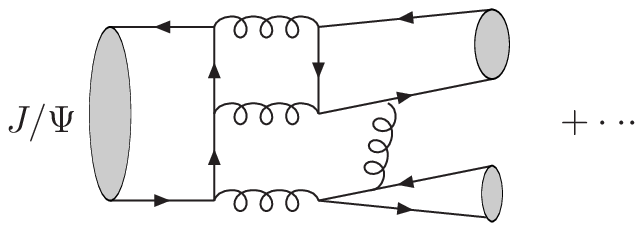}}\\
(a)&&(b)\\
\end{tabular}
\end{center}
\caption{(a) and (b) correspond to the SOZI and DOZI processes for
decays of $J/\psi\rightarrow {two\; light\; mesons}$.}
\label{3-gluon}
\end{figure}

Due to complexity of the loop calculations and non-perturbative
QCD effects which govern the hadronization, accurate estimate of
the contributions from such OZI-suppressed processes to the decay
rates is still missing. Therefore one cannot rule out other
possible mechanisms which also make substantial contributions to
the decay rates of $J/\psi$ and other family members of charmonia.

Meanwhile, in the charm-tau energy regions, some phenomena cannot
be understood in the regular theoretical framework. The $\rho\pi$
puzzle \cite{exp} is the most challenging one. Rosner recently
suggested \cite{Rosner} that this puzzle may be explained by the
$2S-1D$ mixing. There exists an alternative possibility which
might explain the surprisingly small branching ratio of
$\psi(3886)\rightarrow \rho\pi$, i.e. the final state interaction
(FSI) \cite{Anisowich,Li,Cheng}.

Of course the first challenge to theorists is to correctly
evaluate the widths of the exclusive  $J/\psi$ non-leptonic
decays. Besides the OZI suppressed mechanism, there exists another
possible process which may also contribute to the amplitude. That
is the contribution from the hadronic loops.  In fact, except the
phase space for various channels, the mechanism where three gluons
are exchanged, should be flavor-blind and universal for all the
processes.  It is another reason to motivate us to seek for a new
mechanism because a simple analysis indicates that data of
$J/\psi$ decays do not manifest the expected universality.

This picture looks similar to that for the final state interaction
(FSI) \cite{Anisowich,Li}, but essentially different. In the FSI
picture, $J/\psi$ or $\psi(3770)$ decay into two real hadrons and
the two real hadrons re-scatter into another pair of hadrons which
have the same isospin structure, but different identities, via
strong interaction. In that case, the two hadrons, no matter in
the intermediate stage or in the final states, are real and on
their mass shells, therefore the two light hadrons cannot be
charmed mesons such as $D$ or $D^*$ due to constraints of
momentum-energy conservation i.e. $M_{J/\psi}<2M_D (or\;
M_D+M_{D^*}$). One can derive the re-scattering amplitude by
calculating the absorptive (imaginary) part of the so-called
triangle diagrams \cite{Anisowich,Li,Cheng}.

There exists a different contribution. Recently, Suzuki
\cite{Suzuki} indicates that the real part of the triangle where
the two intermediate hadrons in the triangle can be off-shell
virtual ones, does also contribute. Thus to estimate the
corresponding contributions, one needs to calculate the dispersive
(real) part of the triangle.  The diagram corresponding to the
hadronic loops is shown in Fig.2 (there may be more similar
diagrams which  contribute to the processes (see Fig.3)).
\begin{figure}[htb]
\begin{center}
\begin{tabular}{c}
\scalebox{0.9}{\includegraphics{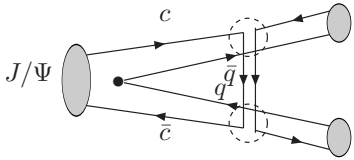}}\\
\end{tabular}
\end{center}
\caption{The hadronic loop diagram where the two hadrons inside
the triangle are time-like, but virtual ones and the exchanged
hadron possessing appropriate quantum numbers are also off-shell.}
\label{haha}
\end{figure}

In fact, such a mechanism should also exist in all the decay modes
of $J/\psi$ and make sizable contributions. In this work, we are
going to evaluate the contribution from the hadronic loops where
the intermediate hadrons are off-mass-shell.

The physical picture is following: $J/\psi$ first dissolves into
two virtual hadrons, then by exchanging an appropriate hadron (i.e
it possesses appropriate charge, flavor, spin and isospin) they
turn into two on-shell real light hadrons which are to be seen by
detector. The two intermediate hadrons are charmed hadrons, which
contains flavor $c$ or $\bar c$, thus the transition
$J/\psi\rightarrow H_1H_2$ where $H_1$ and $H_2$ are the two
charmed hadrons, does not suffer from the OZI suppression. It is
noted that the authors of \cite{Maiani} discuss a possible
mechanism for the relativistic heavy ion collision, where $J/\psi$
absorbs a pion existing in the hot environment to transit into two
charmed mesons and the process may influence the observed $J/\psi$
suppression at RHIC. Even though such process is different from
that under discussion of this work, the pictures have
similarities.

In the practical calculations, the coupling constant of the
effective vertex cannot be obtained by fitting data because such a
process cannot occur due to the limited phase space. Then certain
reasonable symmetries are invoked to get the coupling constants.
We can argue that even though the coupling constant itself is not
accurately determined, the qualitative conclusion is not seriously
influenced.

In contrast with the derivation of the absorptive part of the
triangle, the loop integration seems to be ultraviolet-divergent.
In fact, following the standard procedure, the effective vertices
are obtained from the chiral Lagrangian where all the concerned
hadrons are on-shell, i.e. are real ones. To compensate the
off-shell effects of the hadrons in the triangle, one may
introduce form factors at each vertex. The form factors are
inversely proportional to $(q^2-\Lambda^2)$ where $\Lambda$ is a
phenomenological parameter \cite{Anisowich} which is expressed in
terms of another phenomenological parameter $\alpha$ (see the
following text for details). Because of existence of the form
factors the ultraviolet divergence disappears, namely $\Lambda$
indeed plays a role similar to the cut-offs in the Pauli-Villas
renormalization scheme \cite{Izukson,peskin}.

Picture is as follows, because, so far, we cannot derive the
contribution from the OZI suppressed processes (i.e. via the
three-gluon intermediate stage)  to the total amplitude based on
the quantum field theory yet, we denote such contribution by a
phenomenological parameter $|\mathcal{G}_{S}^{PV}|$ which should
be fixed by fitting data.

Therefore, by the simple picture, there are only two parameters,
i.e. $\alpha$ and $|\mathcal{G}_{S}^{PV}|$. Concretely, we suppose
that both the direct three-gluon process and hadronic loops
contribute to the amplitude altogether, by fitting data of two
distinict channels of $J/\psi\rightarrow PV$ where $P$ and $V$
stand for pseudoscalar and vector mesons respectively, we obtain
the values of $\alpha$ and $|\mathcal{G}_{S}^{PV}|$
simultaneously. Here we choose to study processes
$J/\psi\rightarrow PV$, since there are more data available.  Of
course when we derive $|\mathcal{G}_{S}^{PV}|$ from data, we have
to carefully take care of the phase space of final states.
Applying the obtained $\alpha$ and $|\mathcal{G}_{S}^{PV}|$, we
evaluate the rates of other $J/\psi\rightarrow PV$ channels.

Indeed, our numerical results show that a destructive interference
between the three-gluon process(OZI) and the hadronic loop can
result in rates which are satisfactorily consistent with data. We
will give a detailed discussion in the last section.

This work is organized as follows, after the introduction, in
Sect. II, we formulate the hadronic loop contribution for
$J/\psi\to PV$. In Sect. III, we present our numerical results
along with all the input parameters. Finally, Sect. IV is devoted
to discussion and conclusion.

\section{Formulation}

The effective Lagrangian for $J/\psi$ decaying into light
pseudoscalar and vector mesons is written as \cite{Chung}
\begin{eqnarray}
\mathcal{L}&=&(\mathcal{G}_{S}^{PV}+\mathcal{G}_{H}^{PV})
\epsilon^{\mu\nu\alpha\beta}\epsilon_{\mu}
Tr[\partial_{\nu}{\mathbb{P}}^{\dag}\partial_{\beta}\mathbb{V}_{\alpha}]\nonumber\\&&+
\mathcal{G}_{D}^{PV} \epsilon^{\mu\nu\alpha\beta}\epsilon_{\mu}
Tr[\partial_{\nu}{\mathbb{P}}^{\dag}]Tr[\partial_{\beta}\mathbb{V}_{\alpha}],
\end{eqnarray}
where $\mathcal{G}^{PV}_{_S}$, $\mathcal{G}^{PV}_{_D}$ and
$\mathcal{G}^{PV}_{_H}$ respectively denote the effective
couplings corresponding to single OZI diagram (Fig. \ref{3-gluon}
(a)), double OZI diagram (Fig. \ref{3-gluon} (b)) and hadronic
loop diagram (Fig. \ref{haha}). In this work, we neglect the DOZI
contribution. $\mathbb{P}$ and $\mathbb{V}$ respectively denote
the nonet pseudoscalar and the nonet vector meson matrices
\begin{eqnarray}
\mathbb{P}&=&\left(\begin{array}{ccc}
\frac{\pi^{0}}{\sqrt{2}}+\frac{x_{\eta}\eta+x_{\eta'}\eta'}{\sqrt{2}}&\pi^{+}&K^{+}\\
\pi^{-}&-\frac{\pi^{0}}{\sqrt{2}}+\frac{x_{\eta}\eta+x_{\eta'}\eta'}{\sqrt{2}}&
K^{0}\\
K^- &\bar{K}^{0}&y_{\eta}\eta+y_{\eta'}\eta'
\end{array}\right),\\
\mathbb{V}&=&\left(\begin{array}{ccc}
\frac{\rho^{0}}{\sqrt{2}}+\frac{\omega}{\sqrt{2}}&\rho^{+}&K^{*+}\\
\rho^{-}&-\frac{\rho^{0}}{\sqrt{2}}+\frac{\omega}{\sqrt{2}}&
K^{*0}\\
K^{*-} &\bar{K}^{*0}&\phi
\end{array}\right).
\end{eqnarray}
In the above expression, $x_{\eta,\eta'}$ and $y_{\eta,\eta'}$
describe the $\eta-\eta's$ mixing, and
\begin{eqnarray}
&&x_{\eta}=y_{\eta'}=\frac{\cos\theta-\sqrt{2}\sin\theta}{\sqrt{3}},\;\;\;
x_{\eta'}=-y_{\eta}=\frac{\sin\theta+\sqrt{2}\cos\theta}{\sqrt{3}},
\end{eqnarray}
where $\theta=-19.1^\circ$\cite{etamixing} is the mixing angle of
$\eta$ and $\eta'$.

%\subsection{Hadronic loop contribution}

Taking $J/\psi \to \rho^{0}\pi^{0}$ as an example, we illustrate
the calculations of the contributions from hadronic loops.

\begin{figure}[htb]
\begin{center}
\scalebox{0.7}{\includegraphics{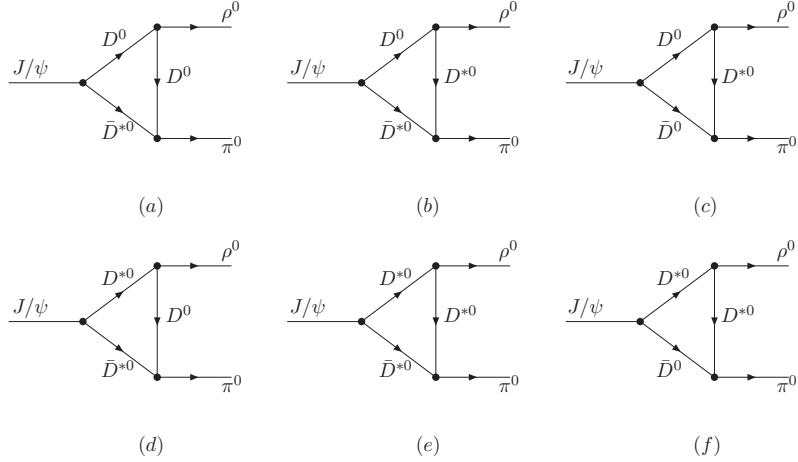}}
\end{center}
\caption{The Feynman diagrams which depict the decays of
$J/\psi\to \rho^0\pi^0$ through $D^{(*)}$ hadronic loops.  }
\label{rho-pi}
\end{figure}

The Feynman diagrams describing the contributions of the hadronic
loops to $J/\psi\to \rho^{0}\pi^{0}$ are depicted in Fig.
\ref{rho-pi}. In the hadronic loops of diagrams (a)-(f) only
$D^{(*)0}$ and $\bar{D}^{(*)0}$ are involved, but of course they
can be simply replaced by $D^{(*)\pm}$ to compose new diagrams.

In the former works \cite{lagrangian-hl,Casalbuoni}, the effective
Lagrangians, which are related to our calculation, are constructed
based on the chiral symmetry and heavy quark symmetry as
\begin{eqnarray}
\mathcal{L}&=&ig_{_{J/\psi
\mathcal{D}{\mathcal{D}}}}[{\mathcal{D}_{i}}{\stackrel{\leftrightarrow}{\partial}}
 _{\mu}\mathcal{D}^{j\dag}]\epsilon^{\mu}+g_{_{J/\psi
\mathcal{D}{\mathcal{D}^{*}}}}\epsilon^{\mu\nu\alpha\beta}\epsilon_{\mu}
\partial_{\nu}{\mathcal{D}_{i}}\partial_{\beta}{\mathcal{D}}^{*j\dag}_{\alpha}\nonumber\\&&
+ig_{_{J/\psi
\mathcal{D^{*}}\mathcal{D^{*}}}}[{\mathcal{D^{*}}}_{i}^{\mu}(\partial_{\mu}
\mathcal{D}^{*j\dag}_{\nu})\epsilon^{\nu}-{\mathcal{D}^{*}}^{j\mu\dag}(\partial_{\mu}
{{\mathcal{D}}^{*}}_{i\nu})\epsilon^{\nu}-(\mathcal{D^{*}}^{\mu}_{i}
{\stackrel{\leftrightarrow}
{\partial}}_{\nu}\mathcal{D}^{*j\dag}_{\mu})\epsilon^{\nu}]\nonumber\\&&
+\bigg\{\frac{1}{2}g_{_{\mathcal{D^{*}}\mathcal{D^{*}}\mathbb{P}}}\varepsilon_
{\mu\nu\alpha\beta}\mathcal{D^* }_{i}^{\mu}
\partial^{\nu}\mathbb{P}^{ij}{\stackrel{\leftrightarrow}{\partial}}^
{\alpha}\mathcal{D^{*}}_{j}^{\beta\dagger}-ig_{_{\mathcal{D^{*}}\mathcal{D}\mathbb{P}}}(\mathcal{D}^{i}\partial^{\mu}\mathbb{P}_{ij}\mathcal{D^{*}}_{\mu}^{j\dagger}
-\mathcal{D^{*}}_{\mu}^{i}\partial
^{\mu}\mathbb{P}_{ij}\mathcal{D}^{j\dagger})\nonumber\\
&&-ig_{_{\mathcal{D}\mathcal{D}\mathbb{V}}}\mathcal{D}_{i}^{\dagger}{\stackrel{\leftrightarrow}{\partial}}
_{\mu}\mathcal{D}^{j}(\mathbb{V}^{\mu})^{i}_{j}
-2f_{_{\mathcal{D^{*}}\mathcal{D}\mathbb{V}}}\varepsilon_{\mu\nu\alpha\beta}(\partial^{\mu}\mathbb{V}^{\nu})
^{i}_{j}(\mathcal{D}_{i}^{\dagger}
{\stackrel{\leftrightarrow}{\partial}}^{\alpha}\mathcal{D^{*}}^{\beta
j}-\mathcal{D^{*}}_{i}^{\beta
\dagger}{\stackrel{\leftrightarrow}{\partial}}^{\alpha}\mathcal{D}^{j})\nonumber\\
&&+ig_{_{\mathcal{D^{*}}\mathcal{D^{*}}\mathbb{V}}}\mathcal{D^{*}}_{i}^{\nu
\dagger}{\stackrel{\leftrightarrow}{\partial}}_{\mu}\mathcal{D^{*}}_{\nu}^{j}(\mathbb{V}^{\mu})^{i}_{j}+
4if_{_{\mathcal{D^{*}}\mathcal{D^{*}}\mathbb{V}}}\mathcal{D^{*}}_{i\mu}^{\dagger}(\partial^{\mu}\mathbb{V}^{\nu}-\partial^{\nu}
\mathbb{V}^{\mu})^{i}_{j}
\mathcal{D^{*}}_{\nu}^{j}\bigg\},\label{lagrangian}
\end{eqnarray}
where $\mathcal{D}$ and $\mathcal{D^*}$ are respectively
pseudoscalar and vector heavy mesons, i.e.
$\mathcal{D^{(*)}}$=(($\bar{D}^{0})^{(*)}$, $(D^{-})^{(*)}$,
$(D_{s}^{-})^{(*)}$). The actual values of the coupling constants
will be given in the following section.

With these Lagrangians, we write out the decay amplitudes of
$J/\psi\rightarrow \rho^{0} \pi^{0}$. For Fig. \ref{rho-pi} (a),
\begin{eqnarray}
\mathcal{M}_{a}&=&\mathcal{A}_{\rho^{0}\pi^{0}}\int\frac{d^4
q}{(2\pi)^4}[i g_{J/\psi DD^*} \varepsilon_{\mu\nu\alpha\beta}
\epsilon_{J/\psi}^{\mu}(ip_{2}^{\nu})(ip_{1}^{\beta})
][ig_{DD\rho}(-p_{1}-q)\cdot
\epsilon_{\rho}][g_{D^{*}D\pi}(ip_{4\sigma})]\nonumber\\
&&\times\frac{i}{p_{1}^2
-m_{D}^2}\frac{i(-g^{\alpha\sigma})}{p_{2}^2
-m_{D^*}^2}\frac{i}{q^2 -m_{D}^2
}\mathcal{F}^2(m_{D},q^2).\label{integral a}
\end{eqnarray}

For saving the space, we collect the concrete expressions of the
amplitudes corresponding to Fig. \ref{rho-pi} (b)-(f) in the
Appendix.

In the amplitudes, $\mathcal{A}_{\rho^{0}\pi^{0}}=1/2$.
$\mathcal{F}(m_{i},q^2)$s represent the form factors which
compensate the off-shell effects of the mesons at the vertices and
are written as \cite{formfactor-pv}
\begin{eqnarray}
\mathcal{F}(m_{i}, q^2)=\frac{\Lambda^2-m_{i}^2}{\Lambda^2-q^2}
\end{eqnarray}
where $\Lambda$ is a phenomenological parameter. It is obvious
that as $q^2\rightarrow 0$ it becomes a number and if $\Lambda\gg
m_{i}$, it turns to be unity. Whereas, as $q^2\rightarrow\infty$,
the form factor approaches to zero. In that situation, the
distance becomes very small, so that the inner structure would
manifest itself and the whole picture of hadron interaction is no
longer valid. At that region the form factor turns to zero and
plays a role to cut off the end effect. The concrete expression of
$\Lambda$ is represented as \cite{HY-Chen}
\begin{eqnarray}
\Lambda(m_{i})=m_{i}+\alpha \Lambda_{QCD},
\end{eqnarray}
where $m_{i}$ denotes the mass of exchanged meson.
$\Lambda_{QCD}=220$ MeV. $\alpha$ is a phenomenological parameter.
In fact, in literature, some other forms for the form factors are
suggested\cite{form-factor}, all of them depend on
phenomenological parameters which are obtained by fitting data,
therefore in the calculations all the forms are somehow equivalent
even though some of them may provide a better convergent behavior
for the triangle loop integration. That is very similar to the
case of the pauli-Villas renormalization scheme
\cite{Izukson,peskin}.

Taking Fig. \ref{rho-pi} (a) as an example, we carry out the Feynman
parameter integration for (\ref{integral a}) and get
\begin{eqnarray}
\mathcal{M}_{(a)}&=&\mathcal{A}_{\rho^{0}\pi^{0}}g_{_{J/\psi DD^*
}}g_{_{DD\rho}}g_{_{D^{*}D\pi}} \int^{1}_{0}dx\int^{1-x}_{0}dy
\bigg[\frac{2}{(4\pi)^2}\log\Big(\frac{\Delta(m_{_{D}},m_{_{D^{*}}},\Lambda(m_{_{D}}))}{
\Delta(m_{_{D}},m_{_{D^{*}}},m_{_{D}})}\Big)\nonumber\\&&-\frac{(\Lambda^2-m_{D}^{2})y}{8\pi^2
\Delta(m_{_{D}},m_{_{D^{*}}},m_{_{D}})}\bigg]
(-2\varepsilon_{\mu\nu\alpha\beta}\epsilon_{J/\psi}^{\mu}\epsilon_{\rho}^{\nu}p_{3}^{\alpha}
p_{4}^{\beta}),\label{pv-1}
\end{eqnarray}
where
\begin{eqnarray*}
\Delta(a,b,c)&=&m_{3}^{2}(1-x-y)^2-2(p_{3}\cdot
p_{4})(1-x-y)x+m_{4}^2
x^2-(m_{3}^{2}-a^{2})(1-x-y)\nonumber\\&&-(m_{4}^{2}-b^{2})x+yc^2,
\end{eqnarray*}
$m_{3}$($m_{4}$), $p_3$($p_4$) are the masses and  four- momenta
of $\rho^{0}$($\pi^{0}$) respectively. Because of the form
factors, the ultraviolet divergence does not exist as expected.

We use the same method to deal with the amplitudes corresponding to
Fig. \ref{rho-pi} (b)-(f) and also put them in the appendix.

To obtain the amplitudes corresponding to hadronic loops which
contain $D^{(*)\pm}$, one only needs to replace the parameters
corresponding to $D^{(*)0}$ and $\bar{D}^{(*)0}$ by that to
$D^{(*)+}$ and $D^{(*)-}$ in the above six expressions
(\ref{pv-1})-(\ref{pv-6}). Thus contribution from the hadronic
loops to the total amplitude can be eventually expressed as
\begin{eqnarray}
\mathcal{M}^{\rho^{0}\pi^{0}}_{H}&=&(\mathcal{M}_{a}+\mathcal{M}_{b}+\mathcal{M}_{c}+\mathcal{M}_{d}
+\mathcal{M}_{e}+\mathcal{M}_{f})\nonumber\\
&&+(\mathcal{M}_{a}+\mathcal{M}_{b}+\mathcal{M}_{c}+\mathcal{M}_{d}
+\mathcal{M}_{e}+\mathcal{M}_{f})|_{m_{D^{(*)0}}\to m_{D^{(*)\pm}}}\nonumber\\
&=&\mathcal{G}^{{\rho^{0}\pi^{0}}}_{_{H}}(\alpha)\varepsilon_{\mu\nu\alpha\beta}\epsilon_{J/\psi}^{\mu}\epsilon_{\rho}^{\nu}p_{3}^{\alpha}
p_{4}^{\beta},
\end{eqnarray}
where $\mathcal{G}^{{\rho^{0}\pi^{0}}}_{_{H}}(\alpha)$ denotes the
effective coupling for $J/\psi\to \rho^{0}\pi^{0}$ induced by the
hadronic loops where
$\mathcal{G}^{{\rho^{0}\pi^{0}}}_{_{H}}(\alpha)$ is a function of
variable $\alpha$.

Similar to $J/\psi\to \rho^{0}\pi^{0}$, we can also obtain
contributions of hadronic loops to $J/\psi\to
K^{*+}K^{-}+\mathrm{c.c.},\;K^{*0}\bar{K}^{0}+\mathrm{c.c.},\;\omega\eta^{(')},
\;\phi\eta^{(')}$. It is noticed that the hadronic loops for
processes $J/\psi\to
K^{*+}K^{-}+\mathrm{c.c.},\;K^{*0}\bar{K}^{0}+\mathrm{c.c.},
\;\phi\eta^{(')}$ include $D_{s}^{(*)(\pm)}$ mesons and in the
calculations, the coefficient $\mathcal{A}_{\rho^{0}\pi^{0}}$ is
replaced by $\mathcal{A}_{K^{*}K}=1$ (or
$\mathcal{A}_{\omega\eta^{(')}}=x_{\eta^{(')}}/2,\;
\mathcal{A}_{\phi\eta^{(')}}=y_{\eta^{(')}}/2$).

We have obtained seven quantities which contain both contributions
of the OZI process and the hadronic loop
\begin{eqnarray}
\mathcal{G}^{\rho^{0}\pi^{0}}&=&\mathcal{G}^{PV}_{S}\beta^{\rho^{0}\pi^{0}}
+\mathcal{G}^{\rho^{0}\pi^{0}}_{H}(\alpha),\label{haha-1}\\
\mathcal{G}^{K^{*+}K^{-}}&=&\mathcal{G}^{PV}_{S}\beta^{K^{*+}K^{-}}
+\mathcal{G}^{K^{*+}K^{-}}_{H}(\alpha),\label{haha-2}\\
%\mathcal{G}^{K^{*0}\bar{K}^{0}}&=&\mathcal{G}^{K^{*0}\bar{K}^{0}}_{S}
%+\mathcal{G}^{K^{*0}\bar{K}^{0}}_{H}(\alpha),\\
\mathcal{G}^{\omega\eta}&=&\mathcal{G}^{PV}_{S}\beta^{\omega\eta}+\mathcal{G}^{\omega\eta}_{D}
+\mathcal{G}^{\omega\eta}_{H}(\alpha),\label{DOZI-1}\\
\mathcal{G}^{\omega\eta'}&=&\mathcal{G}^{PV}_{S}\beta^{\omega\eta'}+\mathcal{G}^{\omega\eta'}_{D}
+\mathcal{G}^{\omega\eta'}_{H}(\alpha),\label{DOZI-2}\\
\mathcal{G}^{\phi\eta}&=&\mathcal{G}^{PV}_{S}\beta^{\phi\eta}+\mathcal{G}^{\phi\eta}_{D}
+\mathcal{G}^{\phi\eta}_{H}(\alpha),\label{DOZI-3}\\
\mathcal{G}^{\phi\eta'}&=&\mathcal{G}^{PV}_{S}\beta^{\phi\eta'}+\mathcal{G}^{\phi\eta'}_{D}
+\mathcal{G}^{\phi\eta'}_{H}(\alpha)\label{DOZI-4},
\end{eqnarray}
where the contributions from the DOZI processes are ignored, i.e.
one can set $\mathcal{G}^{\omega\eta^{(')}}_{D}=
\mathcal{G}^{\phi\eta^{(')}}_{D}=0$. Geometry factors $\beta^{PV}$
satisfy the following relation:
\begin{eqnarray}
\beta^{\rho^{0}\pi^{0}}_{S}:\beta^{K^{*+}K^{-}}_{S}:\beta^{\omega\eta}_{S}:
\beta^{\omega\eta'}_{S}:\beta^{\phi\eta}_{S}:\beta^{\phi\eta'}_{S}=1
:{1}:x_{\eta}:x_{\eta'}:y_{\eta}:y_{\eta'}.
\end{eqnarray}
The SU(3) geometry factors  $\beta^{PV}$ is obtained by taking the
trace $Tr(PV)$ as $J/\psi$ is an SU(3) singlet,  and in fact, it
is just the SU(3) C-G coefficient for the concerned channel.

Thus using eqs. (\ref{haha-1}) and (\ref{haha-2}), the parameter
$\alpha$ is determined. With it, we are able to calculate the
values of $\mathcal{G}^{PV}_{H}$.

\section{Numerical results}

Because of the flavor SU(3) symmetry, it is reasonable to set
$g_{_{J/\psi D_{s}\bar{D}_{s}}}\approx g_{_{J/\psi D\bar{D}}}$.
Based on the naive vector dominance model and using the data of
the branching ratio of $J/\psi\to e^{+}e^{-}$, the authors of Ref.
\cite{Achasov} determined ${g_{_{J/\psi D\bar D}}^2}/{(4\pi)}=5$.
As a consequence of the spin symmetry in the HQET, $g_{_{J/\psi
D_{s}\bar{D}_{s}^{*}}}$ and $g_{_{J/\psi
D_{s}^{*}\bar{D}_{s}^{*}}}$ satisfy the relations: $g_{_{J/\psi
D_{s}\bar{D}_{s}^{*}}}={g_{_{J/\psi
D\bar{D}_{s}}}}/{m_{_{D_{s}}}}$ and $g_{_{J/\psi
D^{*}_{s}\bar{D}_{s}^{*}}}=g_{_{J/\psi D_{s}\bar{D}_{s}}}$
\cite{JPsi-relation}. Other coupling constants related to
$J/\psi\to PV$ include \cite{HY-Chen}:
\begin{eqnarray*}
g_{_{D^{*}D^{*}\pi}}&=&\frac{g_{_{D^{*}D\pi}}}{\sqrt{m_{_{D}}m_{_{D^{*}}}}}=\frac{2g}{f_{\pi}},
\;\;\;g_{_{DDV}}=g_{_{D^{*}D^{*}V}}=\frac{\beta
g_{_{V}}}{\sqrt{2}},\;\;\;f_{_{D^{*}DV}}=\frac{f_{_{D^{*}D^{*}V}}}{m_{_{D^*}}}=\frac{\lambda
g_{_{V}}}{\sqrt{2}},\\
g_{_{D^{*}D_{s}K}}&=&\sqrt{\frac{m_{_{D_{s}}}}{m_{_{D}}}}g_{_{D^{*}D\pi}},\;\;\;
g_{_{D_{s}^{*}DK}}=\sqrt{\frac{m_{_{D_{s}^*}}}{m_{_{D^*}}}}g_{_{D^{*}D\pi}},\;\;\;
g_{_{V}}=\frac{m_{_{\rho}}}{f_{\pi}},
\end{eqnarray*}
where $f_{\pi}=132$ MeV, $g_{_{V}},\;\beta$ and $\lambda$ are
parameters in the effective chiral Lagrangian that describes the
interaction of heavy mesons with low-momentum light vector mesons
\cite{Casalbuoni}. Following  ref. \cite{Isola}, we take $g=0.59$,
$\beta=0.9$ and $\lambda=0.56$ $\mathrm{GeV}^{-1}$\footnote{In
ref.[14] the authors also gave the values of $\lambda$ and $g$ which
are somewhat different from the updated value of $\lambda$ and $g$
presented in [20]. In this work, we use the updated values. Since we
obtain other parameters by fitting data, the final results are not
sensitive to the values of $\lambda$ and $g$}. Meanwhile, the
coupling constants
$g_{D_{s}^{(*)}D_{s}^{(*)}\phi(K^{*})}$($f_{D_{s}^{(*)}D_{s}^{(*)}\phi(K^{*})}$)
are related to $g_{D^{(*)}D^{(*)}\rho}$($f_{D^{(*)}D^{(*)}\rho}$)
and one can expect the relations in the limit of SU(3) symmetry
\cite{HY-Chen}
\begin{eqnarray*}
g_{D_{s}^{(*)}D_{s}^{(*)}\phi}&=&\frac{m_{D_{s}^{(*)}}}{m_{D^{(*)}}}g_{D^{(*)}D^{(*)}\rho},\;\;\;
f_{D_{s}^{(*)}D_{s}^{(*)}\phi}=\frac{m_{D_{s}^{(*)}}}{m_{D^{(*)}}}f_{D^{(*)}D^{(*)}\rho},\\
g_{D^{(*)}D_{s}^{(*)}K^{*}}&=&\sqrt{\frac{m_{D_{s}^{(*)}}}{m_{D^{(*)}}}}g_{D^{(*)}D^{(*)}\rho},\;\;\;
f_{D^{(*)}D_{s}^{(*)}K^{*}}=\sqrt{\frac{m_{D_{s}^{(*)}}}{m_{D^{(*)}}}}f_{D^{(*)}D^{(*)}\rho}.
\end{eqnarray*}

%(a) Determination of the value of $\alpha$

Using eqs. (\ref{haha-1}) and (\ref{haha-2}), we have determined
$\alpha$ and $\mathcal{G}^{PV}_{S}$ as
$$\alpha=0.13,\;\;\;|\mathcal{G}^{PV}_{S}|=4.51.$$

Namely, in our scheme, the hadronic loop contribution to $J/\psi\to
PV$ is theoretically calculated whereas the OZI contribution is
obtained by fitting the data of the two channels. The obtained
$|\mathcal{G}^{PV}_{S}|=4.51$ is supposed to be universal for all
channels of $J/\psi\to PV$ based on SU(3) symmetry. With eqs.
(\ref{DOZI-1})-(\ref{DOZI-4}), we predict the values of
$\mathcal{G}^{PV}$ for all channels of $J/\psi\to PV$. The obtained
results are listed in Table. \ref{pv-table}.

\begin{table}[htb]
\begin{center}
\begin{tabular}{c||cccccccccc} \hline
Decay
mode&$\rho^{0}\pi^{0}$&$K^{*+}K^{-}+c.c.$&$\phi\eta$&$\phi\eta'$&$\omega\eta$&$\omega\eta'$\\\hline\hline
BR$\times10^{-3}$(Experiment)\cite{PDG}&$4.2\pm0.5$&$5.0\pm0.4$&$0.65\pm0.07$&$0.33\pm0.04$
&$1.58\pm0.16$&$0.167\pm0.025$\\\hline $\mathcal{G}^{PV} (10^{-3}$
GeV$^{-1})$&$2.08\pm0.25$&$1.65\pm0.26$&$0.89\pm0.096$&$0.71\pm0.086$&$1.27\pm0.13$&$0.46\pm0.069$\\\hline
$\mathcal{G}^{PV}_{H} (10^{-3}
$GeV$^{-1})$(Theory)&6.59&6.16&3.55&5.13&5.46&4.07\\\hline
$\mathcal{G}^{PV}
(10^{-3}$GeV$^{-1})$(Theory)&2.08(fitting)&1.65(fitting)&0.92&0.62&0.95&0.40\\\hline
BR$\times10^{-3}$(Theory)&4.2(fitting)&5.0(fitting)&0.70&0.25&0.86&0.14\\\hline
\end{tabular}
\end{center}
\caption{The first two modes are well measured, we obtain $\alpha$
and $|\mathcal{G}^{PV}_{S}|=4.51$  by fitting
them.\label{pv-table}}
\end{table}
\vspace{0.6cm}

\section{conclusion and discussion}

From above discussions and numerical results, one has the
following observations.

(1) The contribution from the hadronic loops is sizable and has
the same order of magnitude as that from the SOZI process.

(2) To get reasonable results which are consistent with data, the
interference between the SOZI contribution and that from hadronic
loops is destructive.

(3) We derive the parameter $\alpha$ and the universal
$|\mathcal{G}^{PV}_{S}|$ by fitting data of the first two distinct
modes of $J/\psi$ and present in the first two columns of table 1.
In our picture, there are only two free parameters. The values of
$\mathcal{G}^{PV}$ which we predict in terms of $\alpha$ and
$\mathcal{G}^{PV}_{H}$ are reasonably consistent with the
experimental data.

In this work, we rely on two key assumptions. The first is that
the value $|\mathcal{G}^{PV}_{S}|$ is universal for all channels
of $J/\psi\rightarrow PV$ due to the flavor-blindness of gluon
coupling to quarks. This treatment is similar to that made
\cite{Maiani} for study process $\pi+J/\psi\rightarrow
D^{(*)}+D^{(*)}$ in a hot RHIC environment.

In this work, we re-parameterize $\Lambda$ in terms of $\alpha$
following the ansatz suggested by the authors of  Ref.
\cite{HY-Chen}.  It seems that one may gain some information about
the parameter $\alpha$ from the FSI calculations
\cite{Anisowich,Cheng}, however, it is not really the case. Even
though in both calculations, one deals with the triangle diagrams
which are composed of only hadrons, the situations are different.
For FSI, one calculates the absorptive part where only the
exchanged meson is off-shell, whereas for the calculation of the
dispersive part, all the three mesons in the triangle are
off-shell. Form factors are phenomenologically introduced to
compensate the off-shell effects at each vertex, thus the form
factors should not be the same for calculating the dispersive and
absorptive parts, i.e. even though they have the same expressions,
the involved free parameters may be different. It is noted that in
the FSI case, the value of $\alpha$ in the form factor is expected
to be of order unity \cite{HY-Chen}, but in the case of hadronic
loop the value of $\alpha$ is obviously smaller. In our strategy,
we do not priori set its value, but let data determine it.

As our conclusion, it is obvious that the contribution from the
hadronic loop to $J/\psi\to \;two\;light\; mesons$ should exist.
The question is how large it is compared to that of the SOZI
processes. Namely, if only the SOZI process is responsible for the
decays of $J/\psi$, it is hard to explain the data and as the
contribution from the hadronic loops being added, one can see that
the data are well accommodated within a tolerable range. In our
scenario, there are only two parameters in the whole scenario and
they are determined by fitting data of two channels. With these
two parameters, we evaluate the branching ratios of the other
channels of $J/\psi\rightarrow PV$ and the theoretical predictions
are consistent with data. It indicates that the ansatz is
reasonable and the whole picture makes sense. Indeed, there is
only one channel, i.e. $J/\psi\rightarrow \omega\eta$, for which
the theoretical prediction on the branching ratio is about 0.6 of
the experimental data. Even though there exists such a
discrepancy, for a rough estimate the consistency is not bad at
all and we would like to urge our experimental colleagues to redo
the measurement to get more accurate results. On the other side,
so far the SOZI processes have not well calculated in the
framework of quantum field theory yet, we also need to complete
such a derivation and detect the value which we obtained by
fitting data and assuming a universality for all modes. To be more
accurate, however, further theoretical and experimental efforts
may be needed. This mechanism should exist not only for $J/\psi$
exclusive decays, but also for other members of charmonia.
However, for higher excited states of chamonia, the channels with
final states composed of charmed hadrons are open, and  the
situation becomes more complicated. This picture also applies to
the $\Upsilon$ decays and can be tested by the future experimental
data.

\section*{Acknowledgements} This work is  supported by the National Natural Science Foundation
of China(NNSFC).

\section*{Appendix}

One obtains the decay amplitudes of $J/\psi\to \rho^{0}\pi^{0}$
corresponding to Fig. \ref{rho-pi} (b)-(f).

For Fig. \ref{rho-pi} (b),
\begin{eqnarray}
\mathcal{M}_{b}&=&\mathcal{A}_{\rho^{0}\pi^{0}}\int\frac{d^4
q}{(2\pi)^4}[i g_{J/\psi DD^*} \varepsilon_{\mu\nu\alpha\beta}
\epsilon_{J/\psi}^{\mu}(ip_{2}^{\nu})(ip_{1}^{\beta})
][-{2i}f_{D^*
D\rho}\varepsilon^{\kappa\xi\lambda\tau}(ip_{3}^{\kappa})\epsilon_{\rho}^{\xi}(-ip_{1}^{\lambda}-i
q^{\lambda})]\nonumber\\&&\times
[ig_{D^{*}D^{*}\pi}\varepsilon_{\pi\zeta\delta\omega}(ip_{4}^{\zeta})(-ip_{2}^{\delta})]
\frac{i}{p_{1}^2 -m_{D}^2}\frac{i(-g^{\alpha\pi})}{p_{2}^2
-m_{D^*}^2}\frac{i(-g^{\tau\omega})}{q^2 -m_{D^*}^2
}\mathcal{F}^2(m_{D^*},q^2).
\end{eqnarray}

For Fig. \ref{rho-pi} (c),
\begin{eqnarray}
\mathcal{M}_{c}&=&\mathcal{A}_{\rho^{0}\pi^{0}}\int\frac{d^4
q}{(2\pi)^4}[ig_{J/\psi DD} (p_{1}-p_{2})\cdot
\epsilon_{J/\psi}][-2if_{D^{*}D\rho}\varepsilon^{\kappa\xi\tau\lambda}(ip_{3}^{\kappa})
\epsilon_{\rho}^{\xi}(-ip_{1}^{\lambda}-iq^{\lambda})]\nonumber\\
&&\times[g_{D^{*}D\pi}(ip_{4}^{\mu})]\frac{i}{p_{1}^2
-m_{D}^2}\frac{i}{p_{2}^2 -m_{D}^2}\frac{i(-g_{\mu\tau})}{q^2
-m_{D^*}^2 }\mathcal{F}^2(m_{D^* },q^2).
\end{eqnarray}

For Fig. \ref{rho-pi} (d),
\begin{eqnarray}
\mathcal{M}_{d}&=&\mathcal{A}_{\rho^{0}\pi^{0}}\int\frac{d^4
q}{(2\pi)^4}\{g_{J/\psi D^* D^* }
[ip_{1}^{\nu}\epsilon_{J/\psi}^{\mu}-ip_{2}^{\mu}\epsilon_{J/\psi}^{\nu}-i(p_{2}-p_{1})\cdot
\epsilon_{J/\psi}g^{\mu\nu}]\}\nonumber\\&&\times[2if_{D^*
D\rho}\varepsilon_{\pi\zeta\delta\omega}(ip_{3}^{\pi})\epsilon_{\rho}^{\zeta}(-ip_{1}^{\delta}
-iq^{\delta})][g_{D^* D\pi}(ip_{4}^{\alpha})]\nonumber\\&&\times
\frac{i(-g_{\mu}^{\omega})}{p_{1}^2
-m_{D^*}^2}\frac{i(-g_{\nu\alpha})}{p_{2}^2
-m_{D^*}^2}\frac{i}{q^2 -m_{D}^2 }\mathcal{F}^2(m_{D},q^2).
\end{eqnarray}

For Fig. \ref{rho-pi} (e),
\begin{eqnarray}
\mathcal{M}_{e}&=&\mathcal{A}_{\rho^{0}\pi^{0}}\int\frac{d^4
q}{(2\pi)^4}\{g_{J/\psi D^* D^* }
[ip_{1}^{\nu}\epsilon_{J/\psi}^{\mu}-ip_{2}^{\mu}\epsilon_{J/\psi}^{\nu}-i(p_{2}-p_{1})\cdot
\epsilon_{J/\psi}g^{\mu\nu}]\}\nonumber\\&&\times \{-g_{D^{*}D^*
\rho}(-ip_{1}-iq)\cdot \epsilon_{\rho}g^{\alpha\beta}-4f_{D^* D^*
\rho}(ip_{3}^{\alpha}\epsilon_{\rho}^{\beta}-ip_{3}^{\beta}\epsilon_{\rho}^{\alpha})\}
\nonumber\\&&\times[ig_{D^{*}D^{*}\pi}\varepsilon_{\pi\zeta\delta\omega}(ip_{4}^{\zeta})
(-i p_{2}^{\delta})]\frac{i(-g_{\mu\alpha})}{p_{1}^2
-m_{D^*}^2}\frac{i(-g^{\nu\pi})}{p_{2}^2
-m_{D^*}^2}\frac{i(-g_{\beta}^{\omega})}{q^2 -m_{D^*}^2
}\mathcal{F}^2(m_{D^*},q^2).
\end{eqnarray}

For Fig. \ref{rho-pi} (f),
\begin{eqnarray}
\mathcal{M}_{f}&=&\mathcal{A}_{\rho^{0}\pi^{0}}\int\frac{d^4
q}{(2\pi)^4}[ig_{J/\psi
DD^*}\varepsilon_{\mu\nu\alpha\beta}\epsilon_{J/\psi}^{\mu}(ip_{2}^{\nu})(ip_{1}^{\beta}
)]\nonumber\\&&\times \{-g_{D^{*}D^* \rho}(-ip_{1}-iq)\cdot
\epsilon_{\rho}g^{\delta\tau}-4f_{D^* D^*
\rho}(ip_{3}^{\delta}\epsilon_{\rho}^{\tau}-ip_{3}^{\tau}\epsilon_{\rho}^{\delta})\}
\nonumber\\&&\times[g_{D^*
D\pi}(ip_{4}^{\kappa})]\frac{i(-g^{\alpha}_{\delta})}{p_{1}^2
-m_{D^*}^2}\frac{i}{p_{2}^2 -m_{D}^2}\frac{i(-g_{\tau\kappa})}{q^2
-m_{D^*}^2 }\mathcal{F}^2(m_{D^*},q^2).
\end{eqnarray}

By further calculations, we obtain
\begin{eqnarray}
\mathcal{M}_{(b)}&=&\mathcal{A}_{\rho^{0}\pi^{0}}g_{_{J/\psi
DD^{*}}}2f_{_{DD^{*}\rho}}g_{_{D^{*}D^{*}\pi}}
\int^{1}_{0}dx\int^{1-x}_{0}dy
\bigg\{\Big[\frac{(\Lambda^2-m_{D^*}^{2})y}{16\pi^2
\Delta^2(m_{_{D}},m_{_{D^{*}}},\Lambda(m_{_{D^*}}))}\nonumber
\\
\nonumber&&-\frac{1}{16\pi^2
\Delta(m_{_{D}},m_{_{D^{*}}},m_{_{D^*}})}+\frac{1}{16\pi^2
\Delta(m_{_{D}},m_{_{D^{*}}},\Lambda(m_{_{D^*}}))}\Big]\Big[-2xy^{2}(p_{3}\cdot
p_{4})^2\nonumber\\&&-2x^{2}y(p_{3}\cdot p_{4})^2 +2xy(p_{3}\cdot
p_{4})^2+2xy^2 m_{3}^{2}m_{4}^{2}+2x^2 y m_{3}^{2}m_{4}^{2}- 2xy
m_{3}^{2}m_{4}^{2}\Big]\nonumber\\&&+
\bigg[\frac{2}{(4\pi)^2}\log\Big(\frac{\Delta(m_{_{D}},m_{_{D^{*}}},\Lambda(m_{_{D^*}}))}{
\Delta(m_{_{D}},m_{_{D^{*}}},m_{_{D^*}})}\Big)-\frac{(\Lambda^2-m_{D^*}^{2})y}{8\pi^2
\Delta(m_{_{D}},m_{_{D^{*}}},m_{_{D^*}})}\bigg]\nonumber\\&&\times
\Big[(-6y-2x+1)(p_{3}\cdot p_{4})+2m_{3}^2-2m_{3}^2
x-2m_{3}^{2}y+2y(p_{3}\cdot p_{4})\Big]\bigg\}\nonumber\\&&\times
\varepsilon_{\mu\nu\alpha\beta}\epsilon_{J/\psi}^{\mu}\epsilon_{\rho}^{\nu}p_{3}^{\alpha}
p_{4}^{\beta},\label{pv-2}
\end{eqnarray}

\begin{eqnarray}
\mathcal{M}_{(c)}&=&\mathcal{A}_{\rho^{0}\pi^{0}}g_{_{J/\psi
DD}}f_{_{DD^{*}\rho}}g_{_{DD^{*}\pi}}
\int^{1}_{0}dx\int^{1-x}_{0}dy
\bigg[\frac{2}{(4\pi)^2}\log\Big(\frac{\Delta(m_{_{D}},m_{_{D}},\Lambda(m_{_{D^*}}))}{
\Delta(m_{_{D}},m_{_{D}},m_{_{D^*}})}\Big)\nonumber\\&&
-\frac{(\Lambda^2-m_{D^*}^{2})y}{8\pi^2
\Delta(m_{_{D}},m_{_{D}},m_{_{D^*}})}\bigg]
\Big[-8\varepsilon_{\mu\nu\alpha\beta}\epsilon_{J/\psi}^{\mu}\epsilon_{\rho}^{\nu}p_{3}^{\alpha}
p_{4}^{\beta}\Big],\label{pv-3}
\end{eqnarray}

\begin{eqnarray}
\mathcal{M}_{(d)}&=&\mathcal{A}_{\rho^{0}\pi^{0}}g_{_{J/\psi D^*
D^{*}}}f_{_{DD^{*}\rho}}g_{_{D^{*}D\pi}}
\int^{1}_{0}dx\int^{1-x}_{0}dy
\bigg\{\Big[\frac{(\Lambda^2-m_{D}^{2})y}{16\pi^2
\Delta^2(m_{_{D^{*}}},m_{_{D^*}},\Lambda(m_{_{D}}))}\nonumber
\\
\nonumber&&-\frac{1}{16\pi^2 \Delta(m_{_{D^{*}}},m_{_{D^*
}},m_{_{D}})}+\frac{1}{16\pi^2
\Delta(m_{_{D^*}},m_{_{D^*}},\Lambda(m_{_{D}}))}\Big]\Big[4x^{2}m_{4}^{2}
+4x^2 (p_{3}\cdot p_{4})\nonumber\\&& +4xy(p_{3}\cdot p_{4})\Big]+
12\bigg[\frac{2}{(4\pi)^2}\log\Big(\frac{\Delta(m_{_{D^*}},m_{_{D^{*}}},\Lambda(m_{_{D}}))}{
\Delta(m_{_{D^*}},m_{_{D^{*}}},m_{_{D}})}\Big)-\frac{(\Lambda^2-m_{D}^{2})y}{8\pi^2
\Delta(m_{_{D^*}},m_{_{D^{*}}},m_{_{D}})}\bigg]\bigg\}\nonumber\\&&\times
\varepsilon_{\mu\nu\alpha\beta}\epsilon_{J/\psi}^{\mu}\epsilon_{\rho}^{\nu}p_{3}^{\alpha}
p_{4}^{\beta},\label{pv-4}
\end{eqnarray}

\begin{eqnarray}
\mathcal{M}_{(e)}&=&\mathcal{A}_{\rho^{0}\pi^{0}}g_{_{J/\psi D^*
D^{*}}}g_{_{D^{*}D^{*}\pi}} \int^{1}_{0}dx\int^{1-x}_{0}dy
\bigg\{\Big[\frac{(\Lambda^2-m_{D^{*}}^{2})y}{16\pi^2
\Delta^2(m_{_{D^{*}}},m_{_{D^*}},\Lambda(m_{_{D^{*}}}))}\nonumber
\\
\nonumber&&-\frac{1}{16\pi^2 \Delta(m_{_{D^{*}}},m_{_{D^*
}},m_{_{D^{*}}})}+\frac{1}{16\pi^2
\Delta(m_{_{D^*}},m_{_{D^*}},\Lambda(m_{_{D^{*}}}))}\Big]f_{_{D^{*}D^{*}\rho}}
\Big[4x^{2}m_{3}^{2} \nonumber\\&&+4y^2 m_{3}^2
+4m_{3}^{2}-8xm_{3}^2 -8y m_{3}^{2} +8xym_{3}^{2}+4x^{2}
(p_{3}\cdot p_{4})+4(p_{3}\cdot p_{4})\nonumber\\&&-8x (p_{3}\cdot
p_{4})-4y (p_{3}\cdot p_{4})+4xy(p_{3}\cdot p_{4})\Big]+
\bigg[\frac{2}{(4\pi)^2}\log\Big(\frac{\Delta(m_{_{D^*}},m_{_{D^{*}}},\Lambda(m_{_{D^{*}}}))}{
\Delta(m_{_{D^*}},m_{_{D^{*}}},m_{_{D^{*}}})}\Big)\nonumber\\&&
-\frac{(\Lambda^2-m_{D^{*}}^{2})y}{8\pi^2
\Delta(m_{_{D^*}},m_{_{D^{*}}},m_{_{D^{*}}})}\bigg][16f_{_{D^* D^*
\rho}}-2g_{_{D^* D^* \rho}}]\bigg\}
\varepsilon_{\mu\nu\alpha\beta}\epsilon_{J/\psi}^{\mu}\epsilon_{\rho}^{\nu}p_{3}^{\alpha}
p_{4}^{\beta},\label{pv-5}
\end{eqnarray}

\begin{eqnarray}
\mathcal{M}_{(f)}&=&\mathcal{A}_{\rho^{0}\pi^{0}}g_{_{J/\psi D^*
D}}f_{_{D^{*}D^{*}\rho}}g_{_{DD^{*}\pi}}
\int^{1}_{0}dx\int^{1-x}_{0}dy
\bigg\{\Big[\frac{(\Lambda^2-m_{D^{*}}^{2})y}{16\pi^2
\Delta^2(m_{_{D^{*}}},m_{_{D}},\Lambda(m_{_{D^{*}}}))}\nonumber
\\
\nonumber&&-\frac{1}{16\pi^2 \Delta(m_{_{D^{*}}},m_{_{D
}},m_{_{D^{*}}})}+\frac{1}{16\pi^2
\Delta(m_{_{D^*}},m_{_{D}},\Lambda(m_{_{D^{*}}}))}\Big]
\Big[-4y(p_{3}\cdot p_{4})\Big]\bigg\} \nonumber\\&&\times
\varepsilon_{\mu\nu\alpha\beta}\epsilon_{J/\psi}^{\mu}\epsilon_{\rho}^{\nu}p_{3}^{\alpha}
p_{4}^{\beta}.\label{pv-6}
\end{eqnarray}

\end{document}